\documentclass[aps,pra,twocolumn,superscriptaddress,nobibnotes,a4paper]{revtex4-1}

\usepackage[pdftex]{graphicx}
\usepackage{graphicx}
\usepackage{amsmath}
\usepackage[version=4]{mhchem}
\usepackage{verbatim}
\usepackage{color}

\usepackage[colorlinks=true,allcolors=blue]{hyperref} %hyperrefs

\begin{document}

\title{Nanofabricated tips for device-based scanning tunneling microscopy}\thanks{This work was published in \href{https://doi.org/10.1088/1361-6528/ab1c7f}{Nanotechnology \textbf{30}, 335702 (2019)}.}

\author{Maarten Leeuwenhoek}
\affiliation{Kavli Institute of Nanoscience, Delft University of Technology, Lorentzweg 1, 2628CJ Delft, The Netherlands}
\affiliation{Leiden Institute of Physics, Leiden University, Niels Bohrweg 2, 2333CA Leiden, The Netherlands}

\author{Richard A. Norte}
\affiliation{Kavli Institute of Nanoscience, Delft University of Technology, Lorentzweg 1, 2628CJ Delft, The Netherlands}

\author{Koen M. Bastiaans}
\affiliation{Leiden Institute of Physics, Leiden University, Niels Bohrweg 2, 2333CA Leiden, The Netherlands}

\author{Doohee Cho}
\affiliation{Leiden Institute of Physics, Leiden University, Niels Bohrweg 2, 2333CA Leiden, The Netherlands}

\author{Irene Battisti}
\affiliation{Leiden Institute of Physics, Leiden University, Niels Bohrweg 2, 2333CA Leiden, The Netherlands}

\author{Yaroslav M. Blanter}
\affiliation{Kavli Institute of Nanoscience, Delft University of Technology, Lorentzweg 1, 2628CJ Delft, The Netherlands}

\author{Simon Gr\"{o}blacher}
\email[]{s.groeblacher@tudelft.nl}
\affiliation{Kavli Institute of Nanoscience, Delft University of Technology, Lorentzweg 1, 2628CJ Delft, The Netherlands}

\author{Milan P. Allan}
\email[]{allan@physics.leidenuniv.nl}
\affiliation{Leiden Institute of Physics, Leiden University, Niels Bohrweg 2, 2333CA Leiden, The Netherlands}

\begin{abstract}
We report on the fabrication and performance of a new kind of tip for scanning tunneling microscopy. By fully incorporating a metallic tip on a silicon chip using modern micromachining and nanofabrication techniques, we realize so-called smart tips and show the possibility of device-based STM tips. Contrary to conventional etched metal wire tips, these can be integrated into lithographically defined electrical circuits. We describe a new fabrication method to create a defined apex on a silicon chip and experimentally demonstrate the high performance of the smart tips, both in stability and resolution. In situ tip preparation methods are possible and we verify that they can resolve the herringbone reconstruction and Friedel oscillations on Au(111) surfaces. We further present an overview of possible applications.
\end{abstract}

\maketitle

Scanning tunneling microscopy (STM) is a leading tools for probing electronic and topographic information at the atomic scale~\cite{chen1993introduction}. Since its inception a few decades ago, data quality has dramatically improved by focusing on mechanical stability, tip preparation and lower temperatures~\cite{NIST,Pan1999, Battisti2018}. New possibilities have emerged and greatly extended the range of STM, including quasi-particle interference studies with density of states mapping~\cite{crommie1993imaging,Petersen2000, Fujita2015,Yazdani2015,Fischer2007}, spin-polarized STM~\cite{Wiesendanger2009,Bode2003}, and ultra-low temperature operation~\cite{NIST,Assig2013,VonAllworden2018}.

Here, we introduce a platform for bringing  device-based functionality to STM, with the aim to utilize decades of progress in device engineering for the field of scanning probe. We replace the conventional electrochemically etched, pointy metal wire with an integrated metal tip on a silicon chip. This new platform, which we call smart tip, allows in principle to directly add additional capabilities to a STM tip, including novel spin-sensitivity, local heating, local magnetic fields, local gating, high-frequency compatible coplanar waveguides, qubits, and double-tips. However, it is \emph{a priori} unclear whether  a nanofabricated tip will function for STM measurements, as several challenges arise:\ the stability needs to be below the picometer scale, stringent requirements exist on the shape and sharpness of the freestanding tip, and contamination from fabrication residues need to be absent. In this paper, we demonstrate the feasibility of nanofabricated tips and the the novel smart tip platform. We first discuss our newly developed fabrication procedure and then experimentally show the functionality of these tips in standard STM measurements. 

\begin{figure}[tb!]
	\begin{center}
	\includegraphics[width=1\columnwidth]{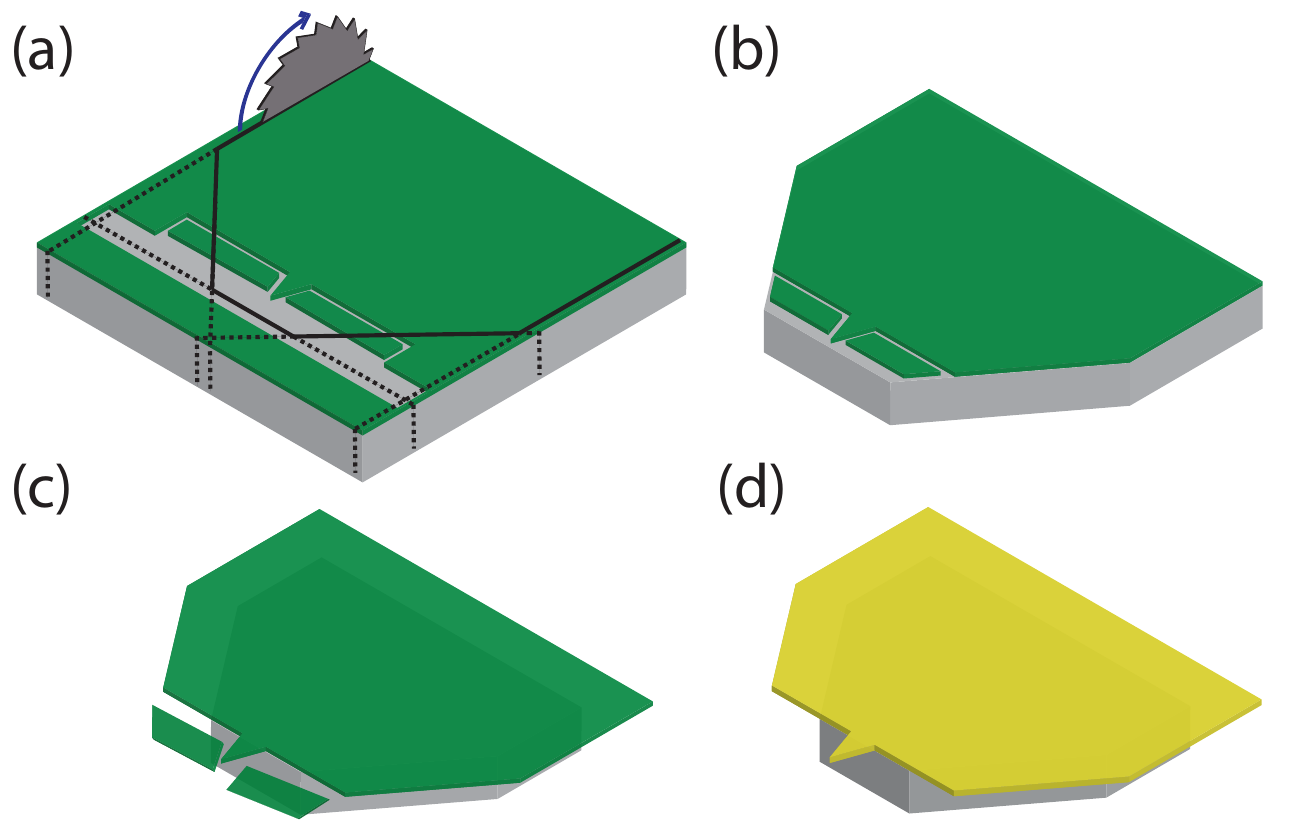}
	\caption{Fabrication procedure for STM smart tips. (a) A layer of SiN (green) covering the Si(100) chip (grey) is patterned through e-beam lithography and plasma etching, creating a large opening and lines that encircle the shields and the tip. (b) The chip is then diced along the black lines. (c) After cleaning, the chip is undercut using a dry release, removing the Si substrate primarily from the sidewalls. The shields are now unsupported and will fall off leaving only the freestanding tip. (d) As a last step, the full chip is covered with a metal, e.g.\ gold, avoiding residues from the processes before.}
	\label{fig:figure1}
	\end{center}
\end{figure}

The challenge in realizing smart tips is to make devices that are fully compatible with conventional STM, yet allow for compatibility with  standard nano- and microfabrication processes. Specifically we need:\ (i) a clear protrusion of the tip relative to the underlying chip, (ii) precise control of the tip shape, and (iii) reliable, reproducible fabrication recipes. To meet these requirements, we developed a new fabrication method using suspended silicon nitride (SiN) tips covered with gold to create on-chip STM smart tips, described below and shown in Fig.~\ref{fig:figure1}.

Using SiN as a base for our suspended STM tips has a number of key advantages. First, it has  a large selectivity to the Si etch we use to suspend the tip, allowing for clear perfusions. Making the tips solely out of metal without underlying SiN would also limit the choice of metal to those compatible with the etch described below. Second, it has a high mechanical stiffness, yielding robust tips  that are resistant to tip treatment, as discussed later. Finally, it allows for a process where the metallic layer is added as a last step. This allows us to avoid any contamination by chemicals such as etchants and resists.

Our process starts with a 500~$\mu$m thick Si(100) chip covered on both sides with a 200~nm thick layer of high stress low-pressure chemical vapor deposition (LPCVD) silicon nitride. We write the initial pattern, consisting of the tip shape and two shields, using electron beam lithography and we then transfer it into the SiN using a reactive-ion \ce{CHF3} etch (cf.\ Fig.~\ref{fig:figure1}(a)). The shields are  slabs of SiN on both sides of the tip, which we  include to minimize the undercut of the Si once the overhang is created:\ the 50~nm lines around the shields reduce the etch rate of the Si significantly compared to a large exposed region without shields. We then clean the chip in a piranha solution to remove all traces of resist and protect it in a new layer of photoresist that can be easily removed later by acetone. In order to bring the tip close to the edge we proceed with dicing the chip along the lines depicted in Fig.~\ref{fig:figure1}(a) resulting in Fig.~\ref{fig:figure1}(b). This step is typically precise to within a few micrometers, leaving a straight sidewall (roughness $<$3~$\mu$m). The residue created by the dicing process is washed away with the removal of the protective photoresist layer. After this cleaning step, we isotropically remove part of the Si substrate using a dry reactive-ion etch (\ce{SF6}). For improved selectivity of the SiN over the Si the chip is cooled to $-50^{\circ}$C. During a typical etch the thickness of the SiN reduces from 200~nm to $\sim$120~nm. The exposed Si sidewalls are removed at a rate of around 4~$\mu$m/min until both the tip and the two shields protrude by about $10-12~\mu$m, causing the shields to fall off (Fig.~\ref{fig:figure1}(c)). The straight sides next to the tip can be made very small or even rounded to avoid accidental touches when aligning to a sample, this would not have been possible using an anisotropic KOH wet etch. The final step in fabricating the STM tip involves depositing a metal on the chip through sputtering to ensure proper coverage of both the top and the side of the SiN tip (Fig.~\ref{fig:figure1}(d)). In this study, we choose to deposit 20~nm of gold using a Leica ACE200 as the tip material; it is relatively straightforward to use other interesting materials for the tip.

\begin{figure*}[tbh!]
	\begin{center}
		\includegraphics[width=1\textwidth]{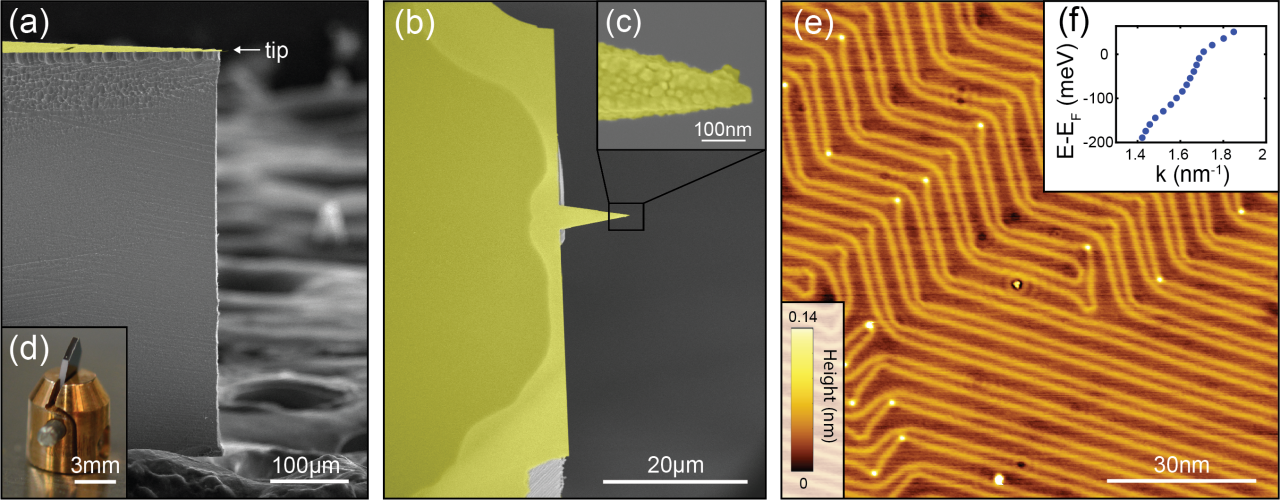}
		\caption{(a-c) Scanning electron microscope (SEM) images of a smart tip from the side (a) and the top (b,c). The freestanding tip made of Au (colored in yellow) covered SiN has a length of 10~$\mu$m and a tip radius of approximately 20~nm. The light yellow area is suspended, while the dark yellow parts are still attached to the underlying Si. (c) The tip radius of this particular device is approximately 20~nm. (d) Photograph of a smart tip mounted in our custom BeCu holder. (e) Topograph of Au(111) surface taken with a fabricated tip at 5.8~K ($V_b=100$~mV, $I=300$~pA). We observe the herringbone surface reconstruction as well as single adatoms (bright dots). In the center a cluster of three adatoms is surrounded by Friedel oscillations. (f) Dispersion of the surface state measured by quasi-particle interference.}
		\label{fig:figure2}
	\end{center} 
\end{figure*}

Images of a typical device are shown in Fig.~\ref{fig:figure2}(a-c). The diameter of the apex of the tip depends on the initial thickness of the SiN, the electron beam spot size and dose, the \ce{SF6} etch time and temperature, and the metal film thickness. However, as can be seen from Fig.~\ref{fig:figure2}(c), the tip diameter is mostly determined by the grain size of the metal film. Our tips achieve radii of a few nanometers, is comparable to specialized commercially available metal wire tips. The overall yield of the fabrication as described in this section is around 80\%.

Our  microfabricated tips rely on and are compatible with existing STM systems and only requires the modification of the tip holder. In this study, we loaded the tips into a modified commercial Unisoku 1500 ultra-high vacuum STM with an operating temperature down to 2.3~K. We customized the BeCu holder for the smart tips, as shown in Fig.~\ref{fig:figure2}(d). The metal covered top side of the chip is placed against the body of the holder and thereby creates a large metal-to-metal surface to ensure good electrical contact (few Ohms) and a clamp pushes against the bottom of the chip. The chip is placed under an angle of 10$^{\circ}$ to avoid contact between the Si sidewall of the chip and the sample. As the metal tip is electrically isolated from the Si by the SiN layer and because Si is highly resistive at low temperatures, tunneling can only originate from the metal tip itself.

Next, we demonstrate the performance of the nanofabricated tips, and show that they are fully compatible with scanning tunnelling microscopy. For this, we use a single smart tip made from gold. After having verified that such tips routinely achieve atomic resolution without any tip preparation in ambient conditions on freshly cleaved graphite, we move to atomically flat Au(111) for more reliable tests. We use an Au(111) film on mica at 5.8~K under UHV conditions. The gold surface is prepared by cycles of Ar ion sputtering (1~kV at $5\times10^{-5}$~mbar) and subsequent annealing at 600$^{\circ}$C for 1~min. We use standard tip cleaning procedures, such as voltage pulsing up to $-$3~V and mechanical annealing~\cite{castellanos2012highly, Tewari2017a}. The latter procedure consists of repeated and controlled indentation of the tip into the surface up to several conductance quanta and leads to a crystalline, atomically sharp tip apex~\cite{Tewari2017a}. All these procedures worked repeatedly with our microfabricated tips.

Figure~\ref{fig:figure2}(e) shows a typical STM image ($99\times99$~nm) of the reconstructed Au(111) surface obtained with the fabricated tip (setup conditions 100~mV, 300~pA). The herringbone reconstruction characteristic to these gold surfaces is resolved in great detail. Several of the kinks of the reconstruction lines house a single adatom identified as bright spots. Around the three contiguous adatoms in the center of the image, we can observe rings that show the waves of electron scattering, also known as Friedel oscillations. Similarly, quasiparticle interference spectroscopy~\cite{Fujita2015} allows us to measure the dispersion of the surface state, shown in Fig.~\ref{fig:figure2}(f). To quantify the lateral resolution of the tip we extract line traces over the atomic step from the topographic data. The width of the step edge is 1.3~nm, matching with values obtained from measurements with a conventional tip on the same system. 

The lateral stability of these tips could be a reason of concern, given their planar nature. Simulations show, however, that the resonance frequency is very high, for both oscillations in the plane and perpendicular to the plane of the tip. Experimentally, we verified that  changing the scan direction in the our of plane direction of the tip does not show any  difference in the performance of the tip. 

We have thus established the functionality of  smart tips and that they are fully compatible with existing (commercial) STM systems. We have tested their performance and demonstrated that common tip preparation techniques such as voltage pulses and mechanical annealing can be applied to improve the tip quality in situ, which is an important aspect when using STM. For the remainder of this paper, we will discuss applications of these tips.

\begin{figure}[htb!]
	\begin{center}
		\includegraphics[width=0.98\columnwidth]{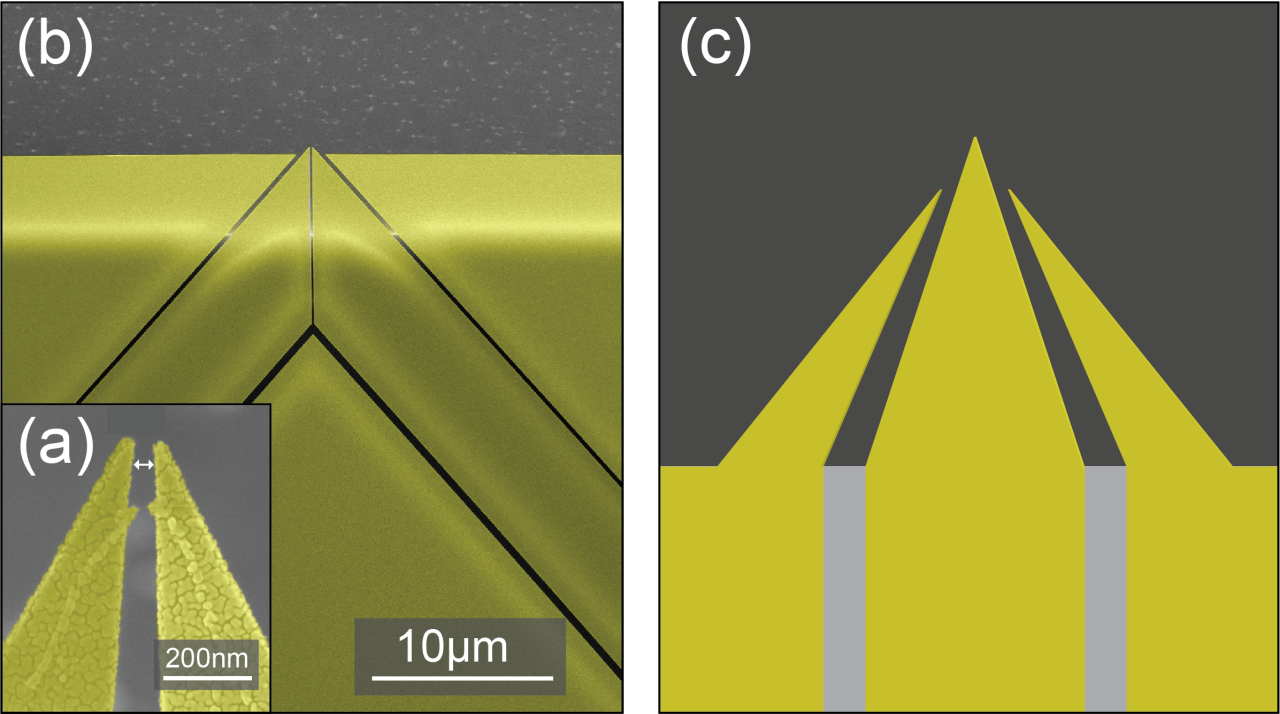}
		\caption{Examples of functionalized smart tips. (a,b) SEM images of a prototype double-tips to measure the Green's functions . The arrow in (a) indicates a spacing of approx.\ 45~nm. The two tips (colored in yellow) are electrically completely separated as shown in (b). (c) Conceptual design of a smart tip with a high-frequency compatible coplanar waveguide, which should be easily adaptable from the double-tip design in (b).}
		\label{fig:figure3}
	\end{center} 
\end{figure}

We would  like to highlight some ways of how smart tips can, in the future, contribute to challenges in condensed matter physics and beyond, with an example shown in Fig.~\ref{fig:figure3}. Such applications beyond standard tips are possible as the fabrication recipes allow for a precise control of the tip shape while achieving atomic resolution, which differentiates it from earlier work that relies on cleaving the chip~\cite{Flohr2012,Siahaan,gurevich2000a}.  A first example is to guide photons  close to the apex of the tip. This can be done for microwave, millimeter and THz waves with coplanar waveguides. There is currently significant research interest in this area~\cite{Baumann417,Donati,paul2017control, THz1, Thz2}, especially with respect to single spin magnetic resonances, charge dynamics, and (shot) noise measurements on the atomic scale. It is however challenging to control the exact microwave radiation power at the junction and to concentrate the power to the area of interest. This can be solved if the tip itself becomes high-frequency compatible. Our smart tip platform allows to directly include coplanar waveguides on the chip, as depicted in Fig.~\ref{fig:figure3}(c). This opens a way for improved spin resonance experiments.

A possible application in condensed matter physics is to measure Green's functions, as suggested previously for experiments with two separate tips~\cite{Niu1995,byers1995probing,settnesPRL,settnesPRB,Ruitenbeek2011, Buttiker1998, Buttiker1999}. Such experiments necessitate very short tip-to-tip distances.  Recently, much progress has been made with arrangements that rely on multiple conventional tips being brought into close proximity, where the distance between tips are limited by the radii of the tips~\cite{bo2000scanning, Jaschinsky2006, roychowdhury201430, Kolmer2017, ge2015development,Hasegawa2007}, but no Green's function measurement has been possible thus far.  Using microfabricated tips has different strengths and weaknesses compared to these approaches. The smart tips can be implemented in compact, ultrastable STM's and are brought into tunneling simultaneously. Microfabrication allows to lithographically define both the distance, as well as the shape of the tips.  We realize a first proof-of-principle demonstration of this new technique by adapting our single-tip into a two-tip pattern, where the trenches between the tips ensure electrical isolation (cf.\ Fig.~\ref{fig:figure3}). This demonstrates that  that we can fabricate tip distances smaller than 50~nm. Further work will concentrate on bringing both tips in tunneling simultaneously, e.g. by mounting them on a piezo to allow for a slight tilt. Mechanical annealing~\cite{mechann2, mechann3, castellanos2012highly, Tewari2017a} can then be used to obtain tips with equal length. {The tips then need to be tested on Au(111) samples to ensure that they have the same properties. }

Further, one could fabricate a gate that is only nanometers away from the probing tip, using a geometry similar to the one shown in Fig.~3b~\cite{gurevich2000a}. Bringing this to the atomic scale allows to image individual donors and their environment in semiconductors and quantum materials. Importantly, there is a large set of quantum materials that are challenging to gate, including high-temperature superconductors. While back- and liquid ion gating had some success~\cite{bollinger2011superconductor}, material issues prevent large tuning or to differentiate between field and chemical gating~\cite{dubuis2016oxygen,deoxygenation}.  We think that this could be especially beneficial in materials with poor electronic screening~\cite{Battisti2017}.

And lastly, the measurement of transport through single molecules, thus far the domain of break junctions~\cite{nitzan2003electron}, can be accessed with smart tips which consist of two tips in close proximity. This has the advantage that one can control where to `contact' the molecules, and one can measure molecules that are not accessible with standard techniques. 

Most importantly, our fabrication recipes allow for easy integration of any standard cleanroom procedure without impeding the STM performance, and therefore new ideas can readily be integrated.

\section{Acknowledgments}
We acknowledge valuable support from the Kavli Nanolab Delft, in particular from C.\ de Boer and M.\ Zuiddam. This project was financially supported by the European Research Council (ERC StG Strong-Q and SpinMelt) and by the Netherlands Organisation for Scientific Research (NWO/OCW), as part of the Frontiers of Nanoscience program, as well as through Vidi grants (680-47-536, 680-47-541).

\clearpage

\end{document}